
\magnification=1200
\hsize=6.5truein \vsize=8.4truein
\parindent 10pt
\parskip 1pt plus 2pt
\baselineskip=17pt

\def\p{\partial} \def\to{\rightarrow}
\def\eqn{\eqno}
\def\implies{\rightarrow}

\overfullrule=0pt
{}~\hfill\vbox{\hbox{TIFR-TH-92-29} \hbox{hepth@xxx/9206016} \hbox{May,
1992}}\break

\vskip .5in

\centerline{\bf MACROSCOPIC CHARGED HETEROTIC STRING}

\vskip .7in

\centerline{Ashoke Sen}

\centerline{\it Tata Institute of Fundamental Research}
\centerline{\it Homi Bhabha Road, Bombay 400005, India}
\centerline{e-mail address: sen@tifrvax.bitnet}

\vskip .7in

\centerline{\bf Abstract}

Classical solutions of equations of motion in low energy effective field
theory, describing fundamental charged heterotic string, are found.
These solutions automatically carry an electric current equal to the
charge per unit length, and hence are accompanied by both, electric and
magnetic fields.
Force between two parallel strings vanish due to cancellation between
electric and magnetic forces, and also between graviton, dilaton, and
antisymmetric tensor field induced forces.
Multi-string solutions describing configuration of parallel strings are
also found.
Finally, the solutions are shown to possess partially broken space-time
supersymmetry.

\def\etwoone{(2.1)}
\def\etwotwo{(2.2)}
\def\etwothree{(2.3)}
\def\etwothreea{(2.4)}
\def\etwofour{(2.5)}
\def\etwofive{(2.6)}
\def\etwosix{(2.7)}
\def\etwosixteenb{(2.8)}
\def\etwoseven{(2.9)}
\def\etwosevena{(2.10)}
\def\etwoeight{(2.11)}
\def\etwoeighta{(2.12)}
\def\etwoeightb{(2.13)}
\def\etwoeightc{(2.14)}
\def\etwonine{(2.15)}
\def\etwoten{(2.16)}
\def\etwoeleven{(2.17)}
\def\etwotwelve{(2.18)}
\def\etwotwelvea{(2.19)}
\def\etwothirteen{(2.20)}
\def\etwofourteen{(2.21)}
\def\etwofifteen{(2.22)}
\def\etwosixteen{(2.23)}
\def\etwosixteena{(2.24)}
\def\etwosixteenc{(2.25)}
\def\etwosixteend{(2.26)}
\def\etwoseventeen{(2.27)}
\def\etwoeighteen{(2.28)}
\def\etwonineteen{(2.29)}
\def\etwotwentyfive{(2.30)}
\def\etwotwentysix{(2.31)}
\def\etwotwentyseven{(2.32)}
\def\etwotwentyeight{(2.33)}
\def\etwotwentynine{(2.34)}

\def\ethreeone{(3.1)}
\def\ethreetwo{(3.2)}
\def\ethreethree{(3.3)}
\def\ethreefour{(3.4)}
\def\ethreefive{(3.5)}
\def\ethreesix{(3.6)}
\def\ethreeseven{(3.7)}
\def\ethreeeight{(3.8)}
\def\ethreenine{(3.9)}
\def\efourone{(4.1)}
\def\efourtwo{(4.2)}
\def\efourthree{(4.3)}
\def\efourfour{(4.4)}
\def\efourfive{(4.5)}
\def\efourfivea{(4.6)}
\def\efoursix{(4.7)}
\def\efiveone{(5.1)}
\def\efivetwo{(5.2)}
\def\efivethree{(5.3)}
\def\efivefour{(5.4)}

\def\RDH{[1]}
\def\RDGHR{[2]}
\def\RWITTEN{[3]}
\def\RHOROST{[4]}
\def\ROTHERSTRING{[5]}
\def\RSTROMSOL{[6]}
\def\ROTHERSOL{[7]}
\def\RMULTI{[8]}
\def\RDUFF{[9]}
\def\RVENEZIANO{[10]}
\def\RSEN{[11]}
\def\RFAWAD{[12]}
\def\RODD{[13]}
\def\RCONFORMAL{[14]}
\def\RSAROJA{[15]}
\def\RFMS{[16]}

\vfill\eject

\centerline{\bf 1. Introduction}

In the study of string theory, much of the recent attention has been
focussed on the construction of classical solutions in string theory.
The motivation for such study is manifold.
First of all, in order to study non-perturbative string theory, one must
include, in addition to the standard Fock space states, the soliton states
in the spectrum.
In particular, some of the underlying `symmetries' of string theory may
become manifest only after including these solitonic states in the
spectrum.
Secondly, since string theory is expected to provide us with a finite
consistent theory of quantum gravity, new insights into phenomena like
black hole evaporation in quantum gravity might be gained by studying
evaporation of black holes in string theory.
Finally, the macroscopic soliton like solutions in string theory might
have been produced in the early universe, and may have observable
consequences and astrophysical implications.

In refs.\RDH\RDGHR\ a specific classical solution in string theory was
constructed which represents fields around a macroscopic heterotic
string\RWITTEN.
It was found that the force between two such parallel strings vanish, and
classical solutions of low energy effective field theory equations of
motion were constructed which represent multi-string solutions.
Finally, it was shown that this solution has partially broken space-time
supersymmetry.
Using space-time supersymmetry, a Bogomol'nyi bound for the mass per unit
length of the string was found, and the solution representing fundamental
string source was shown to saturate this bound.
Later, it was also realized\RHOROST\ that these solutions may be regarded as
the extremal limit of black string solutions in heterotic string theory,
where the horizon approaches the singularity.

Other aspects of macroscopic string-like solutions in string theory have
been discussed in ref.\ROTHERSTRING.
Related work on the construction of
other supersymmetric solutions of string theory equations of motion has
been done in refs.\RSTROMSOL\ROTHERSOL.
Other multi-soliton solutions in string theory have been constructed in
ref.\RMULTI.

In heterotic string theory, besides the string coordinates $X^\mu$, there
are other degrees of freedom.
In particular, it contains 16 internal coordinates $Y^{(I)}$ which are
constrained to be left moving on the world sheet, and are responsible for
giving rise to gauge fields in the spectrum of the theory.
Thus one would expect that if one considers a string source, where the
world-sheet momenta conjugate to the coordinates $Y^{(I)}$ are
non-vanishing, then it would represent a macroscopic string carrying a
finite amount of charge per unit length.
As we shall see later, due to the constraint that the coordinates
$Y^{(I)}$ are left moving, such a string also carries an electric current
equal to the charge per unit length of the
string.\footnote{$^\dagger$}{ Superconducting nature of these strings was
discussed in ref.\RWITTEN.}
In this paper, we shall construct explicit solutions of the classical
equations of motion of the low energy effective field theory, which
represent such charged strings.
The method that we shall be using is the method of twisting\RDUFF-\RODD\
which is
a class of transformations, that act on a classical solution to generate
new inequivalent classical solutions.
In particular, in ref.\RFAWAD\ a class of such transformations were found,
which, acting on an electrically neutral solution, generates an
electrically charged solution.
We shall show that the same transformations, when applied to the solution
of ref.\RDGHR, generate new solutions representing charged macroscopic
string.
Furthermore, most of the features of the charge neutral solution, e.g. no
force between parallel strings, explicit construction of multi-string
solutions, space-time supersymmetry, and fundamental strings as extremal
black strings also hold for the charge carrying solution.

The plan of the paper is as follows.
In sect.2 we shall apply the method of ref.\RFAWAD\ to generate charged
single string solution from the charge neutral solutions of ref.\RDGHR,
and calculate the various charges and effective energy momentum tensor
associated with the solution.
In sect.3 we shall perform the same transformations on the multi-string
solutions of ref.\RDGHR\ and construct a multi-charged-string solution.
This solution, however, is characterized by the fact that all the strings
carry the same amount of charge per unit length.
By slightly modifying this solution we construct multi-string solutions
where different strings carry different amount of charge per unit length.
In  sect.4 we show that the elementary charged string solution corresponds
to the extremal limit of a charged black string solution.
We start from a black string solution carrying no electric or
anti-symmetric tensor gauge field charge, and generate from this a black
string solution carrying both these types of charges by using the
transformation of ref.\RFAWAD.
We then take a specific limit (in which the horizon
approaches the singularity) of this solution and show that the solution
reduces to that
representing a fundamental charged heterotic string.
In sect.5 we discuss supersymmetry transformation properties of the
solution, and show that it is invariant under half of the supersymmetry
transformations of the theory.
We first give a general argument showing that the twisting procedure that
generates charged solutions from the uncharged ones commute with
space-time supersymmetry transformations.
Hence the space-time supersymmetry of the original solution of ref.\RDGHR\
implies that the transformed solution will also be invariant under
space-time supersymmetry.
We then also explicitly verify that for the charged string solution, the
supersymmetry transform of all the fermionic fields in the theory vanish.
We conclude in sect.6 with a summary of our results.

\vskip .5in

\centerline{\bf 2. Single String Solution}

We shall consider heterotic string compactified to $D$ dimensions.
Let us assume that during the process of compactification $p$ of the
original 16 $U(1)$ gauge symmetries remain unbroken.
The interaction between the massless modes of the string theory
and that between the massless modes and the degrees of freedom of the
fundamental
string, are described by the following low energy effective Lagrangian:
$$\eqalign{
S=&-\int d^Dx\sqrt{-G}e^{-\Phi}(-R+{1\over 12} H_{\mu\nu\rho}
H^{\mu\nu\rho} -G^{\mu\nu}\p_\mu\Phi\p_\nu\Phi
+{1\over 8}F^{(I)}_{\mu\nu}F^{(I)\mu\nu})\cr
&-{\mu\over 2}\int d\sigma d\tau(\sqrt{-\gamma} \gamma^{mn}\p_m X^\mu\p_n
X^\nu G_{\mu\nu}(X) +\epsilon^{mn}\p_m X^\mu \p_n X^\nu B_{\mu\nu}(X)\cr
&+C A_\mu^{(I)}(X)\epsilon^{mn}\p_m Y^{(I)}\p_n X^\mu)\cr
}
\eqn\etwoone
$$
Here $G_{\mu\nu}$ is the metric, $R$ is the scalar curvature,
$F^{(I)}_{\mu\nu}=\p_\mu A^{(I)}_\nu-\p_\nu A^{(I)}_\mu$ is the field
strength corresponding to the $U(1)$ gauge fields $A_\mu^{(I)}$ ($1\le I\le
p$), $\Phi$ is the dilaton field,
$$
H_{\mu\nu\rho}=\p_\mu B_{\nu\rho} +{\rm~cyclic~permutations} -
(\Omega_3(A))_{\mu\nu\rho}
\eqn\etwotwo
$$
where $B_{\mu\nu}$ is the antisymmetric tensor gauge field, and,
$$
(\Omega_3(A))_{\mu\nu\rho}={1\over 4} (A^{(I)}_\mu F^{(I)}_{\nu\rho} +
{\rm~cyclic~permutations} )
\eqn\etwothree
$$
is the gauge Chern Simons term.
The metric used here is the $\sigma$-model metric, and is related to the
Einstein metric $G_E$ by the relation
$$
G_{\mu\nu}=exp(2\Phi/(D-2))G_{E\mu\nu}
\eqn\etwothreea
$$
The variables $X^\mu$ denote the coordinates of the string, $\sigma$,
$\tau$ are the world sheet coordinates, and $\gamma_{mn}$ is the world
sheet metric.
$\mu$ is the string tension and
$C$ is a constant $\propto \mu^{-1/2}$ which is adjusted in such a way
that in the low energy effective action involving the massless fields, the
gauge fields appear with the normalization given in eq.\etwoone.
$Y^{(I)}$ denote the internal coordinates of the string responsible for
gauge symmetry, and satisfy,
$$
(\sqrt{-\gamma}\gamma^{mn} -\epsilon^{mn})\p_n Y^{(I)} =0
\eqn\etwofour
$$
where $\epsilon^{mn}$ is the tensor,
$$
\epsilon^{\sigma\tau} = -\epsilon^{\tau\sigma} =1,~~~\epsilon^{\tau\tau}
=\epsilon^{\sigma\sigma}=0
\eqn\etwofive
$$
We shall choose a gauge in which,
$$
(\sqrt{-\gamma})^{-1}\gamma_{mn}= a\eta_{mn} + b P_{mn}
\eqn\etwosix
$$
where $a$ and $b$ are constants to be specified later, $\eta$ is the
Minkowski metric, and,
$$
P=\pmatrix{1 & 1\cr 1& 1}
\eqn\etwosixteenb
$$
The first row/column corresponds to $\sigma$, the second row/column
corresponds to $\tau$.
In this gauge, eq.\etwofour\ takes the form:
$$
(\p_\tau-\p_\sigma) Y^{(I)} =0
\eqn\etwoseven
$$
\vbox{
Note that in writing down the above effective action, we have ignored the
non-Abelian components of the gauge fields, as well as any other massless
fields that might arise during compactification from $10$ to $D$
dimensions.
Similarly, we have ignored the world sheet fermionic degrees of freedom of
the string which carry space-time Lorentz index.
Since we shall look for solutions of the equations of motion where these
degrees of freedom are not excited, this provides a consistent truncation
for our purpose.
We have chosen our mass scale such that $2\kappa^2=1$.

The equations of motion for the fields $G_{\mu\nu}$, $B_{\mu\nu}$,
$\Phi$
and $A_\mu$, derived from the action} \etwoone\ are given
by,\footnote{$^*$}{ Actually, there is a subtlety in deriving the
$A^{(I)}_\nu$ equation.
The left hand side of the $A^{(I)}_\nu$ equation, as derived by varying
the low energy effective action with respect to $A^{(I)}_\nu$, contains an
extra term $-{1\over 2} A^{(I)}_\rho D_\mu(e^{-\Phi}H^{\mu\nu\rho})$,
which, in turn, is proportional to the left hand side of the $B_{\mu\nu}$
equation.
This term is gauge dependent, which is seen by manifest appearance of
$A^{(I)}_\rho$ in the equation.
This apparent inconsistency may be removed as follows.
Since this extra term comes from the variation of the Chern-Simons term
$\Omega_3(A)$ in the effective action, which, in turn, arises from the two
loop $\beta$ function\RCONFORMAL, we need to include the one loop
effective action describing the interaction between {\it world-sheet
fields} and $A_\mu^{(I)}$
in order to get a consistent set of equations of motion.
This effective action is proportional to\RCONFORMAL,
$$
\int d\sigma d\tau A^{(I)}_\mu \p_m X^\mu [\sqrt{-\gamma}\gamma^{mn} -
(\sqrt{-\gamma} \gamma^{mm'} +\epsilon^{mm'}){\p_{m'}\p_{n'}\over \p^2}
(\sqrt{-\gamma} \gamma^{nn'} +\epsilon^{nn'})] A^{(I)}_\nu \p_n X^\nu
$$
If we include this extra term in the expression for the action $S$ given
in eq.\etwoone, and write down the equation of motion for $A^{(I)}_\nu$,
the right hand side of the equation receives an extra term.
This term exactly cancels the gauge non-invariant term on the left hand
side by the $B_{\mu\nu}$ equation of motion if the background satisfies
the condition $\p_\mu A^{(I)}_{\nu}\p_m X^\mu\p_n X^\nu=0$.
Since we shall restrict to backgrounds satisfying this condition, we can
use eq.\etwosevena\ for studying classical solutions of the equations of
motion in the presence of fundamental string background.}
$$\eqalign{
&\sqrt{-G} e^{-\Phi} G^{\mu\xi} G^{\nu\eta} [R_{\mu\nu} -{1\over 2}R
G_{\mu\nu}
+D_\mu D_\nu\Phi -G_{\mu\nu} D^\rho D_\rho\Phi+{1\over 2}G_{\mu\nu} D_\rho
\Phi D^\rho\Phi \cr
&-{1\over 4} (H_{\mu\rho\tau}H_{\nu}^{~~\rho\tau} -{1\over 6}
H_{\rho\tau\sigma} H^{\rho\tau\sigma} G_{\mu\nu})
-{1\over 4} (F_{\mu\rho} F_{\nu}^{~~\rho} -{1\over 4} G_{\mu\nu}
F_{\rho\tau} F^{\rho\tau})]\cr
=& -{\mu\over 2}\int d\sigma d\tau\sqrt{-\gamma} \gamma^{mn} \p_m X^\xi
\p_n X^\eta {\delta^{(D)}(x-X(\sigma,\tau))}\cr
&\sqrt{-G} D_\rho (e^{-\Phi} H^{\mu\nu\rho})=\mu\int d\sigma
d\tau\epsilon^{mn} \p_m
X^\mu \p_n X^\nu {\delta^{(D)}(x-X(\sigma,\tau))}\cr
& R- D_\mu\Phi D^\mu\Phi -{1\over 12} H^2 -{1\over 8} F^2 + 2 D^\mu
D_\mu\Phi =0\cr
& \sqrt{-G}\Big(D_\mu (e^{-\Phi} F^{(I)\mu\nu}) +{1\over 2} e^{-\Phi}
H_{\rho\mu}^{~~~~\nu}
F^{(I)\rho\mu}\Big) \cr
=& C\mu\int d\sigma d\tau\epsilon^{mn}\p_m Y^{(I)} \p_n
X^\nu {\delta^{(D)}(x-X(\sigma,\tau))}\cr
}
\eqn\etwosevena
$$

A solution of these equations of motion
in the presence of a string source of the form:
$$
X^0=\tau, ~~~~X^{D-1}=\sigma, ~~~X^\mu=0 ~{\rm for}~1\le\mu\le D-2, ~~~
Y^{(I)}=0, ~~~~(\sqrt{-\gamma})^{-1}\gamma_{mn}=\eta_{mn}
\eqn\etwoeight
$$
was obtained in ref.\RDGHR.
This solution is given by,
$$\eqalign{
ds^2=&e^E\{ -dt^2 +(dx^{D-1})^2\} +\sum_{i=1}^{D-2} dx^i dx^i\cr
B_{(D-1)t}=&1-e^E,~~~\Phi=E,~~~ A_\mu^{(I)}=0\cr
}
\eqn\etwoeighta
$$
where,
$$
e^{-E}=1 +MG^{(D-2)}(\vec r)
\eqn\etwoeightb
$$
$\vec r$ denotes the $D-2$ dimensional vector $(x^1,\ldots x^{D-2})$, and
$G^{(D-2)}$ is the $D-2$ dimensional Green's function, given by,
$$\eqalign{
G^{(D-2)}(\vec r)=& {1\over (D-4)\omega_{D-3} r^{D-4}}~~~{\rm for}~D>4\cr
=& -{1\over 2\pi}\ln r~~~{\rm for}~D=4\cr
}
\eqn\etwoeightc
$$
where $\omega_{D-3}$ is the volume of a unit $D-3$ sphere.
$M$ is a constant which can be identified to $\mu$ by looking at the
source terms in the equation of motion.

In this case there is no source for the fields $A_\mu^{(I)}$, and as a
result the solution represents fields around a charge neutral heterotic
string.
We shall now discuss the more general case
when the coordinate $Y^{(I)}$ has the form,
$$
\gamma^{\tau m}\p_m Y^{(I)}=p^{(I)}
\eqn\etwonine
$$
where $p^{(I)}$ are a set of constants.
This gives rise to a source for the gauge field, and hence, in general,
represents a charged string.
Note, also, that  due to the constraint given in eq.\etwoseven,
the string acts as a
source for both, the $A_0$ and the $A_{D-1}$ fields.
As a result, the string is accompanied by both, electric and magnetic
fields, and acts as a source of charge density, as well as electric
current.
For convenience of presentation, we shall construct a solution for which
only $p^{(1)}$ is non-zero.
A general solution for which all the $p^{(I)}$'s are non-zero may easily
be found by rotating the final solution in the $p$ dimensional space
spanned by the indices $I$.

The technique that we shall use in constructing a classical solution of
the equations of motion in the presence of such a string source is based
on the method of twisting discussed in refs.\RFAWAD, hence we shall
first briefly review part of the results of ref.\RFAWAD\ which is relevant
for our study.
Let $\{G_{\mu\nu}, B_{\mu\nu},
\Phi, A^{(1)}_\mu\}$ denote a solution of the equations of motion
\etwoseven\
which is independent of $d$ of the $D$ coordinates, including the time
coordinate $t\equiv x^0$.
Let $x^\alpha$ denote these $d$ coordinates, and $x^i$ denote the rest of
the coordinates.
We shall further assume that the components $G_{i\alpha}$ and
$B_{i\alpha}$ vanish, so that both, the metric and the antisymmetric
tensor field, have block diagonal form.
Let us regard $G_{\alpha\beta}$ and $B_{\alpha\beta}$ as $d\times d$
matrices,
and $A^{(1)}_\alpha$ as a $d$ dimensional column vector, with the last
row/column corresponding to the time coordinate.
Let $\eta_{\alpha\beta}$ denote the usual Minkowski metric.
We now define,
$$
K_{\alpha\beta}=  -B_{\alpha\beta} -G_{\alpha\beta} -{1\over 4}
A^{(1)}_\alpha A^{(1)}_\beta
\eqn\etwoten
$$
and the $(2d+1)\times (2d+1)$ matrix,
$$
M=\pmatrix{ (K^T-\eta)G^{-1}(K-\eta) & (K^T-\eta)G^{-1}(K+\eta) &
-(K^T-\eta)G^{-1} A\cr
(K^T+\eta)G^{-1}(K-\eta) & (K^T+\eta)G^{-1}(K+\eta) &
-(K^T+\eta)G^{-1}A\cr
-A^T G^{-1} (K-\eta) & -A^T G^{-1}(K+\eta) & A^TG^{-1}A\cr
}
\eqn\etwoeleven
$$
where $T$ denotes transposition of a matrix.
The result of ref.\RFAWAD\ then says that\break
\noindent $\{G'_{\mu\nu}, B'_{\mu\nu}, \Phi', A^{(1)\prime}_\mu\}$ also
describe a
solution of the classical equations of motion in a region of space where
the right hand sides of eqs.\etwosevena\ are zero (i.e. in the absence of
source terms) if
the primed variables are related to the unprimed ones through the
relations,
$$
M'=\Omega M\Omega^T, ~~\Phi'-\ln\det G' =\Phi-\ln\det G,
{}~~G'_{ij}=G_{ij},~~B'_{ij}=B_{ij}
\eqn\etwotwelve
$$
where $\Omega$ is a matrix of the form
$$
\Omega = \pmatrix{S & \cr & R}
\eqn\etwotwelvea
$$
$S$ and $R$ being arbitrary $O(d-1, 1)$ and $O(d,1)$ matrices which
preserve the
matrices $\eta$ and $\pmatrix{\eta &\cr & 1\cr}$ respectively.

The solution given in  eq.\etwoeighta\ is independent of the coordinates
$t$ and $x^{D-1}$, and is block diagonal.
Thus we may generate new solutions from it by making the transformations
given in eq.\etwotwelve\ with $d=2$.
We choose,
$$
\Omega =\pmatrix{I_3 &&\cr & \cosh\alpha &\sinh\alpha\cr & \sinh\alpha
&\cosh\alpha\cr }
\eqn\etwothirteen
$$
where $I_n$ is the $n \times n$ identity matrix, and $\alpha$ is an
arbitrary number.
Applying this transformation on the solution given in eq.\etwoeighta\ we
get a new solution given by,
$$\eqalign{
ds^{2} =& {1\over \cosh^2{\alpha\over 2} e^{-E} -\sinh^2{\alpha\over
2}} (-dt^2 +(dx^{D-1})^2) \cr
& +{\sinh^2{\alpha\over 2} (e^{-E}-1)\over
(\cosh^2{\alpha\over 2} e^{-E} -\sinh^2{\alpha\over
2})^2} (dt+dx^{D-1})^2
+\sum_{i=1}^{D-2} dx^i dx^i\cr\cr
B_{(D-1)t} =& {\cosh^2{\alpha\over 2}(e^{-E}-1)\over \cosh^2{\alpha\over
2} e^{-E} -\sinh^2{\alpha\over
2}}\cr\cr
A^{(1)}_{D-1} = &A^{(1)}_t ={\sinh\alpha (e^{-E}-1)\over
\cosh^2{\alpha\over 2} e^{-E} -\sinh^2{\alpha\over
2}}\cr\cr
\Phi =& -\ln(\cosh^2{\alpha\over 2}e^{-E} -\sinh^2{\alpha\over 2})\cr
}
\eqn\etwofourteen
$$
Using eq.\etwoeightb\ the solution may be written as,
$$\eqalign{
ds^2 =& {1\over 1+NG^{(D-2)}(\vec r)} (-dt^2 +(dx^{D-1})^2)
+{q^2 G^{(D-2)}(\vec r)\over 4N ( 1+NG^{(D-2)}(\vec r))^2}
(dt+dx^{D-1})^2\cr
&+\sum_{i=1}^{D-2} dx^i dx^i\cr\cr
B_{(D-1)t} =& {NG^{(D-2)}(\vec r)\over  1+NG^{(D-2)}(\vec r)}\cr\cr
A^{(1)}_{D-1} =& A^{(1)}_t = {q G^{(D-2)}(\vec r)\over  1+NG^{(D-2)}(\vec
r) }\cr\cr
\Phi= &-\ln( 1+NG^{(D-2)}(\vec r))\cr\cr
}
\eqn\etwofifteen
$$
where
$$
N=M\cosh^2{\alpha\over 2},~~~~~q=M\sinh\alpha
\eqn\etwosixteen
$$

The solution is not invariant under a boost in the $x^{D-1}$ direction,
but such a boost corresponds to changing the parameter $q$.
Note that the solution becomes singular as $\vec r\to 0$, showing that
there are possible source terms at $\vec r=0$.
These source terms may be calculated by explicitly evaluating the $\delta$
function singularities on the left hand side of eqs.\etwosevena, and can
be
shown to be consistent with the following string configuration:
$$
X^{0}=\tau, ~~~~ X^{D-1}=\sigma, ~~~~
\gamma^{\tau m}\p_m Y^{(1)}=p^{(1)}
\eqn\etwosixteena
$$
provided we make the identification,
$$
\mu = N, ~~~~ p^{(1)} = - {q\over \mu C}
\eqn\etwosixteenc
$$
Here $\gamma$ is the world sheet metric induced by the target space
metric, and satisfies,
$$
(\sqrt{-\gamma})^{-1}\gamma_{mn} =\eta_{mn} +{q^2\over 4N^2} P_{mn}
\eqn\etwosixteend
$$

The various field strength tensors may be calculated from this solution
and we get the following results,
$$\eqalign{
F^{(1)}_{r (D-1)} =& F^{(1)}_{rt} = - {q\over r^{D-3}\omega_{D-3} (
1+NG^{(D-2)}(\vec r))^2}\cr
H_{r (D-1) t} =& -{N\over  r^{D-3}\omega_{D-3} (
1+NG^{(D-2)}(\vec r))^2}\cr
}
\eqn\etwoseventeen
$$
Note that $\Omega_3(A)$ vanishes everywhere for the specific solution we
have constructed.
For $D>4$ the Einstein metric defined in eq.\etwothreea\ is asymptotically
flat, and the electric charge $Q$, the electric current $J$ and the
axionic charge
$Z$ associated with the solution may be defined in terms of the asymptotic
behavior of the field strengths in the $r\to\infty$ limit as follows:
$$\eqalign{
{1\over 2\sqrt 2} F^{(1)}_{rt} &\simeq -{Q\over r^{D-3}\omega_{D-3}}\cr
{1\over 2\sqrt 2} F^{(1)}_{r (D-1)} &\simeq -{J\over
r^{D-3}\omega_{D-3}}\cr
H_{r (D-1) t} &\simeq -{Z\over r^{D-3}\omega_{D-3}}\cr
}
\eqn\etwoeighteen
$$
(The factor of $2\sqrt 2$ in the definition of the electric charge and
electric current has been introduced so that our normalization matches
that of refs.\RHOROST.)
Eqs.\etwoseventeen\ and \etwoeighteen\ give,
$$
Q={q\over 2\sqrt 2}, ~~~~ J={q\over 2\sqrt 2}, ~~~~ Z=N
\eqn\etwonineteen
$$
Thus the string carries an electric current equal to its electric charge
per unit length, as expected from the general arguments given before.

For $D>4$, the energy momentum tensor associated with the solution may be
calculated
by the procedure used in ref.\RDGHR.
We first compute the Einstein metric using eq.\etwothreea, and define,
$$
h_{\mu\nu} = G_{E\mu\nu}-\eta_{\mu\nu}
\eqn\etwotwentyfive
$$
{}From this we can compute the linearised Ricci tensor as,
$$
R^{(1)}_{\mu\nu} ={1\over 2} \Big( {\p^2 h^\rho_\mu\over \p x^\rho\p x^\nu}
+{\p^2 h^\rho_\nu\over \p x^\rho\p x^\mu}-{\p^2 h^\rho_\rho\over \p
x^\mu\p x^\nu} -{\p^2 h_{\mu\nu}\over \p x^\rho\p x_\rho}\Big)
\eqn\etwotwentysix
$$
where the indices are raised or lowered by the metric $\eta_{\mu\nu}$.
The total energy momentum tensor $\Theta_{\mu\nu}$ is then defined
as\RDGHR,
$$
\Theta_{\mu\nu} = 2(R^{(1)}_{\mu\nu} -{1\over 2}
R^{(1)\rho}_{~~~\rho}\eta_{\mu\nu} )
\eqn\etwotwentyseven
$$
The effective two dimensional energy momentum tensor $T_{\alpha\beta}$
associated
with the string may then be obtained as,
$$\eqalign{
T_{\alpha\beta} =&\int d^{D-2} x\Theta_{\alpha\beta}(x)\cr
=& \int_{S^{(D-3)}}\Big[-{\p h_{\alpha\beta}\over\p x^l}-\eta_{\alpha\beta}
\big\{{\p h_{kl}\over\p x^k}+{\p h_{tt}\over\p x^l}-{\p h_{(D-1)(D-1)}
\over\p x^l}-{\p h_{kk}\over\p x^l}\big\}\Big] n^l r^{D-3}
d\Omega_{D-3}\cr
}
\eqn\etwotwentyeight
$$
where we have performed an integration by parts.
Here $S^{(D-3)}$ denotes the surface at $r=\infty$, and $n^i$ is a vector
normal to the surface.
Evaluating the left hand side of eq.\etwotwentyeight\ for the present
solution, we get,
$$
T_{\alpha\beta} = N (-\eta_{\alpha\beta} + {q^2\over 4N^2}
P_{\alpha\beta} )
\eqn\etwotwentynine
$$
where the matrix $P$ has been defined in eq.\etwosixteenb.

\vskip .5in

\centerline{\bf 3. Multi-string Solutions}

In this section we shall construct explicit solutions which represent
multiple parallel strings at rest.
In order to show that such a solution is at all possible, we shall first
show that in the field of the single string given in eq.\etwofifteen, a
test string parallel to the original string does not encounter any force.
The test string is given in the following gauge:
$$
X^0=\tau, ~~~~X^{(D-1)}=\sigma, ~~~~(\sqrt{-\gamma})^{-1} \gamma_{mn}=
a\eta_{mn} +b P_{mn}
\eqn\ethreeone
$$
where $a$ and $b$ are two constants whose value we shall not need for our
analysis.
The equations of motion for the transverse coordinates $X^i$ derived from
the action given in eq.\etwoone\ then takes the form:
$$\eqalign{
\sqrt{-\gamma}\gamma^{mn}\p_m\p_n X^i =& - \sqrt{-\gamma}\gamma^{mn}
\Gamma^i_{\nu\rho}\p_m X^\nu\p_n
X^\rho  +{1\over 2} H^i_{~~\nu\rho}\p_m X^\nu
\p_n X^\rho \epsilon^{mn}\cr
&+ CG^{ij}F^{(I)}_{j\mu} \epsilon^{mn}\p_m Y^{(I)} \p_n X^\mu\cr
}
\eqn\ethreetwo
$$
We now start with the following configuration of the test string at a
given time,
$$
\p_\tau X^i = \p_\sigma X^i =0,~~~~ (\p_\tau -\p_\sigma) Y^{(I)} = 0
\eqn\ethreethree
$$
Eq.\ethreetwo\ then gives,
$$
{\p^2 X^i\over \p\tau^2} =0
\eqn\ethreefour
$$
showing that the test string does not encounter a force in the transverse
direction.

Let us now turn to the problem of explicitly constructing the multi-string
solution.
As a first step, we take the multi string solution of ref.\RDGHR\ and
transform it by the transformation given in eq.\etwotwelve.
The final result has the form given in eq.\etwofourteen, except that
$e^{-E}$ is now given by,
$$
e^{-E} = 1 +\sum_l M G^{(D-2)}(\vec r -\vec r_l)
\eqn\ethreefive
$$
where $\vec r_l$ are the locations of the strings.
Substituting this into eq.\etwofourteen\ we get the following multi-string
solution:
$$\eqalign{
ds^2 =& {1\over 1+N\sum_l G^{(D-2)}(\vec r-\vec r_l)} (-dt^2
+(dx^{D-1})^2)\cr
&+{q^2 \sum_l G^{(D-2)}(\vec r-\vec r_l)\over 4N ( 1+N\sum_l G^{(D-2)}(\vec
r-\vec r_l))^2}
(dt+dx^{D-1})^2
+\sum_{i=1}^{D-2} dx^i dx^i\cr
B_{(D-1)t} =& {N\sum_l G^{(D-2)}(\vec r-\vec r_l)\over
1+N\sum_lG^{(D-2)}(\vec r-\vec r_l)}\cr
A^{(1)}_{D-1} =& A^{(1)}_t = {q \sum_l G^{(D-2)}(\vec r-\vec r_l)\over
1+N\sum_l G^{(D-2)}(\vec
r-\vec r_l) }\cr
\Phi= &-\ln( 1+N\sum_l G^{(D-2)}(\vec r-\vec r_l))\cr
}
\eqn\ethreesix
$$
Although this represents a charged multi-string solution,
we would like to construct more general multi-string solutions where the
different strings carry independent $U(1)$ charges $Q_l^{(I)}$.
Examining eq.\ethreesix, we take the following ansatz for such a
multi-string solution:
$$\eqalign{
ds^2 =& {1\over 1+N\sum_l G^{(D-2)}(\vec r-\vec r_l)} (-dt^2
+(dx^{D-1})^2)\cr
& + g(\vec r)
(dt+dx^{D-1})^2
+\sum_{i=1}^{D-2} dx^i dx^i\cr
B_{(D-1)t} =& {N\sum_l G^{(D-2)}(\vec r-\vec r_l)\over
1+N\sum_lG^{(D-2)}(\vec r-\vec r_l)}\cr
A^{(I)}_{D-1} =& A^{(I)}_t = { \sum_l q_l^{(I)} G^{(D-2)}(\vec r-\vec
r_l)\over
1+N\sum_l G^{(D-2)}(\vec
r-\vec r_l) }\cr
\Phi= &-\ln( 1+N\sum_l G^{(D-2)}(\vec r-\vec r_l))\cr
}
\eqn\ethreeseven
$$
where $g(\vec r)$ is a function to be determined.
With this ansatz, the equations of motion for $B_{\mu\nu}$, $\Phi$ and
$A_\mu^{(I)}$ are satisfied identically.
The equation of  motion for the metric gives rise to the following
differential equation for the function $g(\vec r)$:
$$\eqalign{
&\sum_{k=1}^{D-2} \p_k\p_k\Big\{ g(\vec r) \big(1+\sum_l N G^{(D-2)}(\vec
r-\vec
r_l)\big)\Big\}\cr
=&-{1\over 2} \big(1+\sum_l NG^{(D-2)}(\vec r -\vec r_l)\big)
\sum_{I=1}^{16}\sum_{k=1}^{D-2} \bigg(\p_k\bigg({\sum_l q^{(I)}_l
G^{(D-2)}(\vec r-\vec r_l)\over 1+\sum_l N G^{(D-2)}(\vec r -\vec
r_l)}\bigg)\bigg)^2\cr
}
\eqn\ethreeeight
$$
A solution to the above equation is given by,
$$
g(\vec r) ={1\over 4} \bigg[{\sum_{l,I}(q^{(I)}_l)^2 G^{(D-2)}(\vec r-\vec
r_l)\over N\big(1 +\sum_l NG^{(D-2)}(\vec r-\vec r_l)\big)}
-{\sum_I\big(\sum_l q_l^{(I)} G^{(D-2)}(\vec r-\vec r_l)\big)^2 \over
\big(1 +\sum_l NG^{(D-2)}(\vec r-\vec r_l)\big)^2}\bigg]
\eqn\ethreenine
$$
Eqs.\ethreeseven\ and \ethreenine\ gives the general multi-string
solution.

\vskip .5in

\centerline{\bf 4. Fundamental Strings as Extremal Black Holes}

In ref.\RHOROST\ it was shown that the charge neutral fundamental string
solutions of ref.\RDGHR\ can be regarded as extremal limit of black
string solution.
In this section we shall show that even the charged fundamental string can
be regarded as extremal limit of charged black string solutions.
In order to construct the charged black string solutions whose extremal
limit are these fundamental strings, we start with the black string
solution in $D(>4)$ dimensions without any electric, magnetic, or
antisymmetric tensor gauge field charge.
The solution is given by,
$$\eqalign{
ds^2 =& -(1 - {r_+^{D-4}\over r^{D-4}}) dt^2 +{dr^2\over 1- {r_+^{D-4}\over
r^{D-4}}} + r^2 d\Omega_{D-3}^2 + (dx^{D-1})^2\cr
\Phi =& 0, ~~~~ A_\mu^{(I)}=0, ~~~~ B_{\mu\nu}=0\cr
}
\eqn\efourone
$$
where $d\Omega_{D-3}$ denotes the line element of a  $D-3$ sphere.
The solution is independent of the coordinates $t$ and $x^{D-1}$.
We now perform the transformation given in eq.\etwotwelve\ with the
following choice of the matrix $\Omega$:
$$
\Omega = \pmatrix{ I_2 &&& \cr & \cosh\alpha_2 & \sinh\alpha_2 &\cr &
\sinh\alpha_2 & \cosh\alpha_2 & \cr &&& 1\cr}
\pmatrix{ I_3 &&\cr & \cosh\alpha_1 & \sinh\alpha_1\cr &\sinh\alpha_1 &
\cosh\alpha_1\cr }
\eqn\efourtwo
$$
The transformed solution in the $D=5$ case was explicitly worked out in
ref.\RFAWAD.
The solution for general $D$ can be obtained from the results of
ref.\RFAWAD\ in a straightforward manner, and we get,
$$\eqalign{
ds^2 =& -{1\over 4 (r^{D-4} - r_0^{D-4})^2} (4 r^{D-4} (r^{D-4} - r_+^{D-4})
-\beta^2 r_+^{2D-8}) dt^2 + (dx^{D-1})^2\cr
& +\beta {r_+^{D-4}\over r^{D-4} - r_0^{D-4}} dx^{D-1} dt + {dr^2 \over 1-
{r_+^{D-4}\over r^{D-4}}} + r^2 d\Omega_{D-3}^2\cr
B_{t(D-1)} =&\beta {r_+^{D-4}\over 2 (r^{D-4} - r_0^{D-4})}\cr
A^{(1)}_t =&\gamma {r_+^{D-4}\over r^{D-4}- r_0^{D-4}}\cr
A^{(1)}_\mu=& 0~~~~~{\rm for}~1\le\mu\le D-1\cr
\Phi =& -\ln (1-{r_0^{D-4}\over r^{D-4}})\cr
}
\eqn\efourthree
$$
where,
$$\eqalign{
\gamma =& \sinh\alpha_1\cr
\beta =& \cosh\alpha_1 \sinh\alpha_2\cr
(r_0)^{D-4} =& {1\over 2} (r_+)^{D-4} (1-\sqrt{1+\beta^2+\gamma^2})\cr
}
\eqn\efourfour
$$
The above solution describes a black string with horizon at $r=r_+$, and
carry charges associated with the antisymmetric tensor gauge field
$B_{\mu\nu}$ and $U(1)$ gauge field $A_\mu^{(1)}$ proportional to $\beta
(r_+)^{D-4}$ and $\gamma (r_+)^{D-4}$ respectively.
$r_+\to 0$ corresponds to the extremal limit of the black string, since in
this limit the horizon approaches the singular point.

We shall now show that the extremal limit of the black hole corresponds to
the fundamental black string solution that we have constructed in the
previous section.
This is done in two stages.
First we boost the solution in the $x^{D-1}$ direction by an angle $\eta$.
Then we take the limit $r_+\to 0$, $\beta\to -\infty$, $\gamma\to\infty$,
$\eta\to\infty$, with the following combinations kept
fixed:
$$
C_1 ={1\over 2}|\beta| (r_+)^{D-4}, ~~~~~ C_2= \gamma\cosh\eta
(r_+)^{D-4},
{}~~~~~\gamma^2 =2|\beta|
\eqn\efourfive
$$
The resulting solution is of the form:
$$\eqalign{
ds^2 =& {1\over 1+{C_1\over r^{D-4}}} (-dt^2 + (dx^{D-1})^2) +{C_2^2\over
4C_1} {1\over r^{D-4} (1 +{C_1\over r^{D-4}})^2} (dt + dx^{D-1})^2\cr
&+ (dr^2 + r^2 d\Omega_{D-3}^2)\cr
B_{(D-1)t} =& {C_1\over r^{D-4}+ C_1}\cr
A^{(1)}_t =& A^{(1)}_{D-1} = {C_2\over r^{D-4}+ C_1}\cr
\Phi =& -\ln (1 +{C_1\over r^{D-4}})\cr
}
\eqn\efourfivea
$$
This is identical to the solution given in eq.\etwofifteen\ with the
identification,
$$
C_1 = {N\over (D-4)\omega_{D-3}}, ~~~~~ C_2 ={q\over (D-4)\omega_{D-3}}
\eqn\efoursix
$$

\vskip .5in

\centerline{\bf 5. Space-time Supersymmetry}

The original solution constructed in ref.\RDGHR\ had a partially broken
space-time supersymmetry.
Thus it is natural to ask if the solution given in eq.\etwofifteen\ also
possesses such a supersymmetry.
We shall first give a general argument showing that the transformation
given in eqs.\etwotwelve, \etwothirteen\ commutes with the space-time
supersymmetry
transformation, hence if the original transformation is space-time
supersymmetric, so must be the transformed solution.
We shall then verify explicitly that the transformed solution is indeed
space-time supersymmetric.

Let us first recall the string field theoretic origin of the `symmetry'
transformation given in eq.\etwotwelve, with $\Omega$ given in
eq.\etwothirteen\RFAWAD.
In string field theory, a general off-shell string field configuration
corresponds to a state in the combined matter-ghost superconformal field
theory.
Restricting to string field configurations that are independent of the
coordinates $x^0$ and $x^{D-1}$, and has only abelian gauge field
configurations, correspond to restricting to conformal field theory states
carrying zero momentum in the $x^0$, $x^{D-1}$ and $Y^{(I)}$ directions.
In particular, the dependence of the corresponding vertex operators on
the world sheet fields
$X^0$ and $Y^{(1)}$ is through powers of $\p^n X^0$, $\bar\p^m X^0$, and
$\bar\p^l Y^{(1)}$.
The correlation functions involving these vertex operators are invariant
under the transformation,
$$
\p^m X^0\to \p^m X^0, ~~~~ \pmatrix{\bar \p^m X^0\cr \bar \p^m Y^{(1)}}
\to
\pmatrix{ \cosh\alpha & \sinh\alpha\cr \sinh\alpha & \cosh\alpha\cr}
\pmatrix{\bar \p^m X^0\cr \bar \p^m Y^{(1)}} ~~~~\forall~ m
\eqn\efiveone
$$
Since the interaction vertices of string field theory are constructed in
terms of the correlation functions in the conformal field theory, this
symmetry of the conformal field theory correlation functions gives rise to
a `symmetry' of string field theory action for this restricted class of
field configutrations.
This is the origin of the `symmetry' given in eqs.\etwotwelve,
\etwothirteen\ in the
effective field theory.

Since so far we do not have a consistent closed heterotic string field
theory (although we do have such a theory involving only the Neveu-Schwarz
states\RSAROJA) we do not know precisely how the space-time supersymmetry
operator
will look like in the string field theory.
However, from the general analysis\RFMS\ it is clear that space-time
supersymmetry will act only on the holomorphic part of the vertex
operators representing a general off-shell string field configuration.
Since the symmetry transformation given in eq.\efiveone\ acts on
the anti-holomorphic part of the vertex operators, these
two symmetry transformations commute.
As a result, \efiveone, or, equivalently, \etwotwelve\ with $\Omega$ given
in eq.\etwothirteen, will transform a
supersymmetric solution to a supersymmetric solution.

To verify explicitly that the solution given in eq.\etwofifteen\
(and also in eqs.\ethreesix, \ethreeseven) has
partial space-time supersymmetry, we need to show the existence of a
supersymmetry transformation parameter $\xi$ such that variations of the
gravitino field $\psi_\mu$, dilatino field $\lambda$, and the gaugino
field $\chi^{(I)}$ vanish.
The corresponding constraints on $\xi$ can be expressed as\RSTROMSOL\
$$\eqalign{
\delta\psi_\mu =0 &\implies D_\mu\xi -{1\over 8}
H_{\mu\nu\rho}\Gamma^{\nu\rho} \xi =0\cr
\delta\lambda =0 &\implies (\not\p\Phi)\xi -{1\over 6}H_{\mu\nu\rho}
\Gamma^{\mu\nu\rho} \xi =0\cr
\delta\chi^{(I)}=0 &\implies F^{(I)}_{\mu\nu}\Gamma^{\mu\nu}\xi =0\cr
}
\eqn\efivetwo
$$
where $\Gamma^{\mu_1\ldots \mu_n}$ denote the antisymmetrized product of
the $\gamma$-matrices.
(Note that the fields and the supersymmetry transformation parameters in
our convention are related to those used in ref.\RDGHR\ by a set of field
redefinitions.)
These equations may be satisfied by choosing,
$$
\xi = e^{\Phi\over 4}\xi_0
\eqn\efivethree
$$
where $\xi_0$ is a constant spinor, satisfying,
$$
\sqrt{-\det\pmatrix{ G_{tt} & G_{t(D-1)}\cr G_{t(D-1)} &
G_{(D-1)(D-1)}\cr}} (\Gamma^t\Gamma^{(D-1)} -\Gamma^{(D-1)}\Gamma^t) \xi_0
=2 \xi_0
\eqn\efivefour
$$
This shows that the solutions are invariant under half of the space-time
supersymmetry generators, since the supersymmetry, instead of being
generated by an arbitrary spinor, is generated by a spinor satisfying the
constraint given in eq.\efivefour.

\vskip .5in

\centerline{\bf 6. Conclusion}

In this paper we have constructed solutions of the classical equations of
motion of low energy effective field theory describing single and multiple
charged heterotic strings parallel to each other.
These solutions are characterized by the novel feature that a charged
string always carries a current equal to its charge per unit length, and
hence is accompanied by both electric and magnetic fields.
We have also shown that the charged string solution may be regarded as the
extremal limit of a charged black string solution.
Finally, these solutions were shown to be invariant under half of the
space-time supersymmetry generators of the theory.
It will be interesting to investigate the cosmological and astrophysical
implications of these charge and current carrying strings.

\vskip .5in

\centerline{\bf References}

\item{\RDH}
A. Dabholkar and J. Harvey, Phys. Rev. Lett. {\bf 63} (1989) 719.

\item{\RDGHR}
A. Dabholkar, G. Gibbons, J. Harvey and F.R. Ruiz, Nucl. Phys. {\bf B340}
(1990) 33.

\item{\RWITTEN}
E. Witten, Phys. Lett. {\bf B153} (1985) 243.

\item{\RHOROST}
G. Horowitz and A. Strominger, Nucl. Phys. {\bf B360} (1991) 197.

\item{\ROTHERSTRING}
S. Naculich, Nucl. Phys. {\bf B296} (1988) 837;
J. Harvey and S. Naculich, Phys. Lett. {\bf B217} (1989) 231;
B.R. Greene, A. Shapere, C. Vafa and S.T. Yau, Nucl. Phys. {\bf b337}
(1990) 1;
S.J. Rey, Phys. Rev. {\bf D43} (1991) 526.

\item{\RSTROMSOL}
A. Strominger, Nucl. Phys. {\bf B343} (1990) 343.

\item{\ROTHERSOL}
G. Gibbons and C. Hull, Phys. Lett. {\bf 109B} (1982) 190;
C. Callan, J. Harvey and A. Strominger, Nucl. Phys. {\bf B359} (1991) 611,
{\bf B367} (1991 60, preprint EFI-91-66;
J. Harvey and A. Strominger, Phys. Rev. Lett. {\bf 66} (1991) 549;
C. Callan, preprint PUPT-1278;
R. Kallosh, preprint SU-ITP-92-1;
R. Kallosh, A. Linde, T. Ortin, A. Peet and A. Van Proeyen, SU-ITP-92-13.

\item{\RMULTI}
R. Khuri, preprints CTP-TAMU-33/92, 35/92, 38/92, 44/92.

\item{\RDUFF}
S. Ferrara, J. Scherk and B. Zumino, Nucl. Phys. {\bf B121} (1977) 393;
E. Cremmer, J. Scherk and S. Ferrara, Phys. Lett. {\bf B68} (1977) 234;
{\bf B74} (1978) 61;
E. Cremmer and J. Scherk, Nucl. Phys. {\bf B127} (1977) 259;
E. Cremmer and B. Julia, Nucl. Phys.{\bf B159} (1979) 141;
M. De Roo, Nucl. Phys. {\bf B255} (1985) 515; Phys. Lett. {\bf B156}
(1985) 331;
E. Bergshoef, I.G. Koh and E. Sezgin, Phys. Lett. {\bf B155} (1985) 331;
M. De Roo and P. Wagemans, Nucl. Phys. {\bf B262} (1985) 646;
L. Castellani, A. Ceresole, S. Ferrara, R. D'Auria, P. Fre and E. Maina,
Nucl. Phys. {\bf B268} (1986) 317; Phys. Lett. {\bf B161} (1985) 91;
S. Cecotti, S. Ferrara and L. Girardello, Nucl. Phys. {\bf B308} (1988)
436;
M. Duff, Nucl. Phys. {\bf B335} (1990) 610.

\item{\RVENEZIANO}
G. Veneziano, Phys. Lett. {\bf B265} (1991) 287;
K. Meissner and G. Veneziano, Phys. Lett. {\bf B267} (1991) 33; Mod. Phys.
Lett. {\bf A6} (1991) 3397;
M. Gasperini, J. Maharana and G. Veneziano, Phys. Lett. {\bf B272} (1991)
277;
M. Gasperini and G. Veneziano, preprint CERN-TH-6321-91.

\item{\RSEN}
A. Sen, Phys. Lett. {\bf B271} (1991) 295; {\bf 274} (1991) 34; preprint
TIFR-TH-92-20.

\item{\RFAWAD}
S.F. Hassan and A. Sen, Nucl. Phys. {\bf B375} (1992) 103.

\item{\RODD}
J. Horne, G. Horowitz and A. Steif, Phys. Rev. Lett. {\bf 68}, 568 (1992);
M. Rocek and E. Verlinde, preprint IASSNS-HEP-91-68;
P. Horava, preprint EFI-91-57;
A. Giveon and M. Rocek, preprint IASSNS-HEP-91-84;
S. Kar, S. Khastagir and A. Kumar, preprint IP-BBSR-91-51;
J. Panvel, preprint LTH 282.

\item{\RCONFORMAL}
A. Sen, Phys. Rev. Lett. {\bf 55} (1985) 1846;
C.G. Callan, D. Friedan, E. Martinec and M. Perry, Nucl. Phys. {\bf B262}
(1985) 593.

\item{\RSAROJA}
R. Saroja and A. Sen, preprint TIFR-TH-92-14 (to appear in Phys. Lett. B).

\item{\RFMS}
D. Friedan, E. Martinec and S. Shenker, Nucl. Phys. {\bf B271} (1986) 93.

\end